%% file: main.tex
\documentclass[%
 reprint,
superscriptaddress,
 amsmath,amssymb,
 aps,
 prl,
floatfix,
]{revtex4-2}
\usepackage{csquotes}
\usepackage[utf8]{inputenc}
\usepackage{graphicx}
\usepackage{dcolumn}
\usepackage{bm}
\usepackage{xcolor}
\usepackage{microtype}
\usepackage[
    separate-uncertainty = true,
    ]{siunitx}

\begin{document}

\title{Dynamics of electronic phase separation at the laser-induced insulator-metal transition in (La$_{0.6}$Pr$_{0.4}$)$_{0.7}$Ca$_{0.3}$MnO$_3$}

\author{Tim Titze}
\affiliation{Georg-August-Universität Göttingen, I. Physikalisches Institut, 37077 G\"ottingen, Germany\looseness=-1}
\author{Maximilian Staabs}
\affiliation{Georg-August-Universität Göttingen, I. Physikalisches Institut, 37077 G\"ottingen, Germany\looseness=-1}
\author{Pia Henning}
\affiliation{Georg-August-Universität Göttingen, I. Physikalisches Institut, 37077 G\"ottingen, Germany\looseness=-1}
\author{Karen Stroh}
\affiliation{Georg-August-Universität Göttingen, I. Physikalisches Institut, 37077 G\"ottingen, Germany\looseness=-1}
\author{Stefan Mathias}
\affiliation{Georg-August-Universität Göttingen, I. Physikalisches Institut, 37077 G\"ottingen, Germany\looseness=-1}
\affiliation{Georg-August-Universität Göttingen, International Center for Advanced Studies of Energy Conversion (ICASEC), 37077 G\"ottingen, Germany}
\author{Vasily Moshnyaga}
\affiliation{Georg-August-Universität Göttingen, I. Physikalisches Institut, 37077 G\"ottingen, Germany\looseness=-1}
\author{Daniel Steil}
\email[]{dsteil@gwdg.de}
\affiliation{Georg-August-Universität Göttingen, I. Physikalisches Institut, 37077 G\"ottingen, Germany\looseness=-1}

\begin{abstract}
Ultrafast optical excitations allow creating new metastable and hidden states in quantum materials. However, the fundamental material properties required to support new emergent order are largely unknown. Here we show for two colossal magnetoresistive (CMR) manganites that electronic phase separation (EPS) strongly favors non-thermal behavior and exploit this to stabilize an optically-induced conducting state. Our results shed light on the role of EPS in optical control of CMR manganites and provide guidance for the design of materials that can exhibit non-equilibrium states of matter.
\end{abstract}

\maketitle
Perovskite manganites like Pr$_{1-x}$Ca$_x$MnO$_3$ and La$_{1-x}$Ca$_x$MnO$_3$ are well known to be susceptible to photoinduced insulator-to-metal transitions, as demonstrated by numerous studies in the past~\cite{Fiebig1998, Fiebig2000, Polli2007, Beaud2009, Li2013, Beaud2014, Zhang2016, Esposito2018, Esposito2018b, Koshihara2022}. The underlying mechanism behind laser-induced metallization in manganites is the annihilation of Jahn-Teller (JT) distortions or polarons. This is achieved by photo-ionization of the Mn$^{3+}$ state in an pulsed laser experiment, but can also be realized via static control parameters, like hydrostatic pressure or electric and magnetic fields~\cite{Asamitsu1997, Tokura1999, Basov2011}. However, a question that is becoming increasingly relevant to all of these studies is the extent to which pulsed optical excitations can be exploited to create long-lived metastable or hidden phases in these materials, and what material properties are required to support such new emergent order.

Here, we  study the role of electronic phase separation (EPS) on the dynamics of optically-induced phase transitions in such materials. We compare optical reflectivity and electrical resistance dynamics in strain-free thin films of optimally doped manganites La$_{0.7}$Ca$_{0.3}$MnO$_3$ (LCMO) and (La$_{0.6}$Pr$_{0.4}$)$_{0.7}$Ca$_{0.3}$MnO$_3$ (LPCMO) after pulsed laser excitation for different temperatures across their metal-insulator transition (MIT). The two systems exhibit highly contrasting behavior: While the LCMO film shows dominantly thermally driven dynamics, the LPCMO film exhibits complex and strongly nonthermal photo-induced reflectivity and resistance dynamics close to the MIT on all investigated timescales $t=1-1000$\,ns. We explain how different EPS causes the strongly divergent dynamics in LCMO and LPCMO.

Crucial to our study, LCMO and LPCMO both exhibit the colossal magnetoresistive (CMR) effect~\cite{Chahara1993, Helmolt1993, Uehara1999}, i.e., a strong reduction of electrical resistance in an applied magnetic field, which is also viewed as a field-induced insulator-to-metal transition. An important difference between LCMO and LPCMO is, however, that LPCMO undergoes a 1$^{st}$ order ferromagnet/paramagnet (FM/PM) phase transition (PT) around $T_C\approx 190-200$\,K, which is accompanied by electronic/structural phase separation at a mesoscopic scale in bulk~\cite{Uehara1999} or at the nanoscale in thin films~\cite{Moshnyaga2014}. In contrast, LCMO exhibits a close to 2$^{nd}$ order continuous phase transition at significantly higher $T_C\approx 260-270$\,K without pronounced phase separation~\cite{Uehara1999}. The difference in EPS of LCMO and LPCMO originates from a tiny change in the average radius of the A-site cation, $\langle r_A \rangle$. Namely, a smaller $\langle r_A\rangle$ in LPCMO compared to LCMO results in an enhancement of electron-phonon coupling and stabilization of static JT distortions, thus, favoring a localization tendency of charge carriers and weakening of ferromagnetism~\cite{Mathur2003}. The resulting, more distorted orthorhombic structure of LPCMO hosts a very large amount of correlated JT polarons (CP); $n_{CP}\approx 1.5$\,\% of the total charge carrier concentration~\cite{Moshnyaga2014}, pushing $T_C$ below the charge ordering (CO) temperature, $T_{CO}=220-230$\,K. These CPs play a decisive role in the MIT as they actuate an AFM coupling of neighboring FM nano-domains at $T \approx T_C$, resulting in a pronounced increase of electrical resistance in LPCMO~\cite{Moshnyaga2014}. In LCMO with a less distorted structure and higher $T_C>T_{CO}$ the amount of CPs is very small, $n_{CP}\approx 10^{-3}-10^{-4}$\,\%~\cite{Moshnyaga2009}, the phase separation with AFM-coupled FM nanodomains is practically missing and the observed CMR values are much smaller than those in LPCMO. This difference between LMCO with hardly any EPS, and LCPMO with distinct nanoscale EPS makes these otherwise similar materials ideal for our investigations.

\begin{figure}[t]
     \centering
     \includegraphics[width=1\columnwidth]{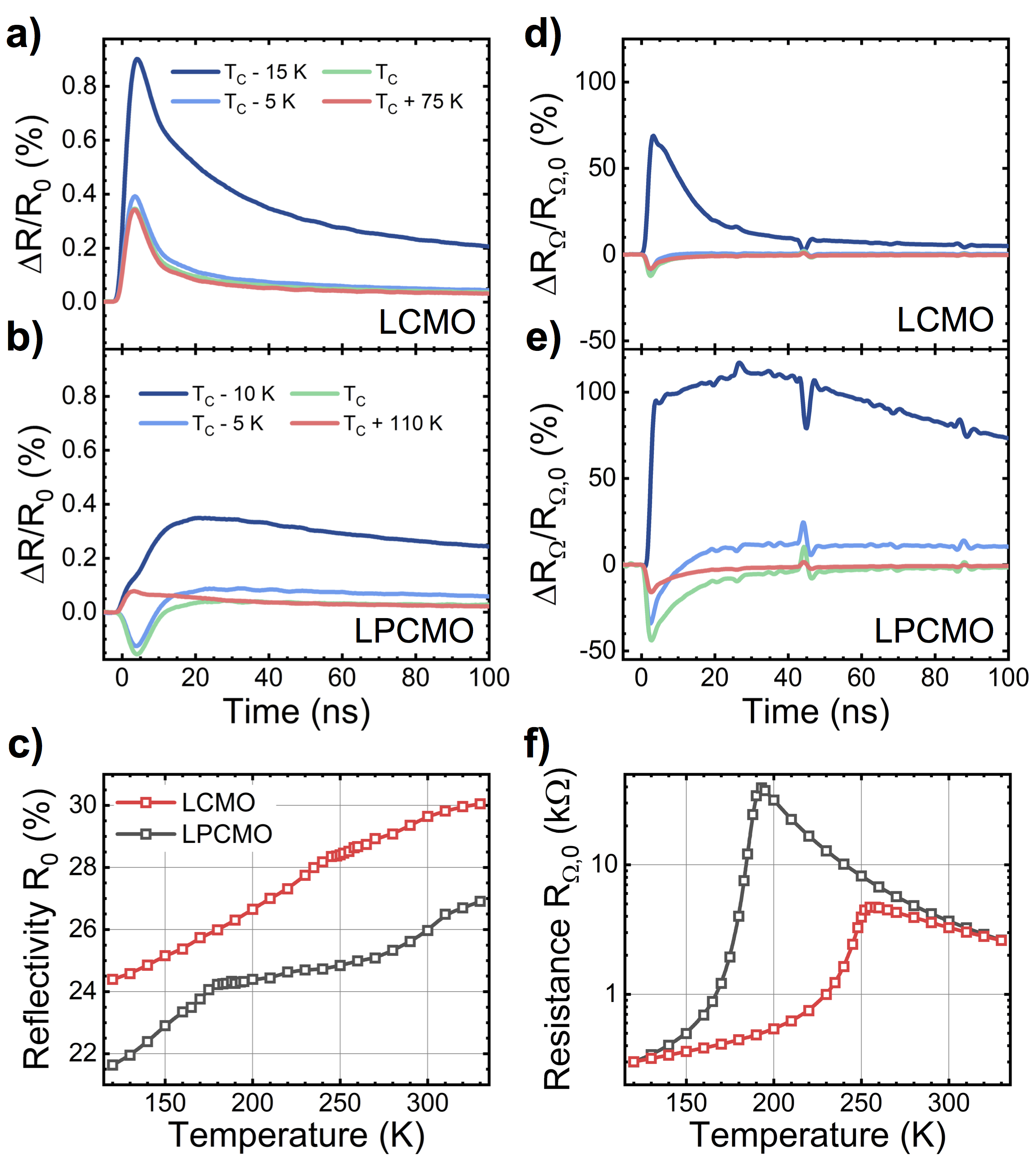}
     \caption{Normalized transient reflectivity for temperatures below (blue), close to (green) and above (red) $T_C$ for a)~LCMO and b)~LPCMO films for $F=3.5$\,mJ/cm$^2$ incident fluence; c)~shows the static $T$-dependent changes of reflectivity in LCMO (red) and LPCMO (black) before time zero. Normalized transient resistance for the same temperatures for d)~LCMO and e)~LPCMO. The small additional dips and peaks on the transients are due to a slight impedance mismatch of the measurement electronics and do not have a further physical meaning; f)~shows the temperature dependences of electrical resistance of LCMO (red) and LPCMO (black) measured in equilibrium.}
     \label{fig:figure1}
\end{figure}

Fig.~\ref{fig:figure1} shows the normalized reflectivity change $[R(t)-\nolinebreak[4] R_0]/R_0=\Delta R(t)/R_0$ of a) LCMO ($T_C\ = 255$\,K) and b) LPCMO ($T_C\ = 190$\,K) films for selected temperatures measured below (blue), in the vicinity of (green) and above (red) the corresponding $T_C$ for 1.25\,ns pulsed laser excitation at $\lambda=515$\,nm. In the case of LCMO, a fast increase in reflectivity can be seen within $t<5$\,ns for all temperatures, followed by a slower decrease within few tens of ns. This is in clear contrast to the LPCMO film, where the transient reflectivity $\Delta R(t)/R_0$ exhibits a sign change from positive ($T\gg T_C$ and $T< T_C$) to negative for temperatures $T \approx T_C$ at early times. Furthermore, the overall reflectivity dynamics in LPCMO significantly slows down within the FM phase with initial state recovery times of up to a microsecond. The full set of T-dependent data are presented in Fig.~\ref{fig:S1} and Fig.~\ref{fig:S2} in the supplemental material (SM).

Assuming that the optical excitation can be described as a purely thermal perturbation on the studied ns timescales, the measurements of static reflectivity for both samples at $\lambda=515$\,nm (see Fig.~\ref{fig:figure1}c) show that a temperature increase should lead to an increase of reflectivity for $T=120-330$\,K. For LCMO, the transient reflectivity data fits qualitatively well to such a simple thermal model describing (i) laser heating of the thin film sample due to the pump excitation and (ii) subsequent thermal relaxation. For LPCMO, the same model obviously fails for temperatures close to $T_C$, since the observed decrease of reflectivity cannot stem from plain sample heating.

To disclose the origin of the non-thermal response in LPCMO we have carried out measurements of transient electrical resistance, see Figs.~\ref{fig:figure1}d~and~\ref{fig:figure1}e. In the case of LCMO, the resistance initially increases for $t<5$\,ns in the FM metallic state ($T<T_{C}$) and decreases in the PM insulating state for $T>T_{C}$. A relatively fast relaxation back to the ground state occurs on the $\approx 10$\,ns time scale. Considering the static temperature-dependent resistance $R_{\Omega}(T)$ (see Fig.~\ref{fig:figure1}f), the transient resistance measurements indicate a dominantly thermal response of LCMO to laser excitation. Indeed, as expected for $T < T_C$ within the FM metallic state, $\Delta R_{\Omega}(t)/dT>0$ and for $T > T_C$ in the PM insulating state $\Delta R_{\Omega}(t)/dT<0$. This observation fully agrees with the results of transient reflectivity (see Figs.~\ref{fig:figure1}a~and~b). In contrast, LPCMO shows a significant nonthermal decrease of the resistance slightly below $T_C$, i.e., $\Delta R_{\Omega}(t)/dT<0$ within the overall FM metallic phase, where static $R_{\Omega}(T)$ would yield $\Delta R_{\Omega}(t)/dT>0$, similar to the observations made in transient reflectivity.

\begin{figure}[b]
     \centering
     \includegraphics[width=1\columnwidth]{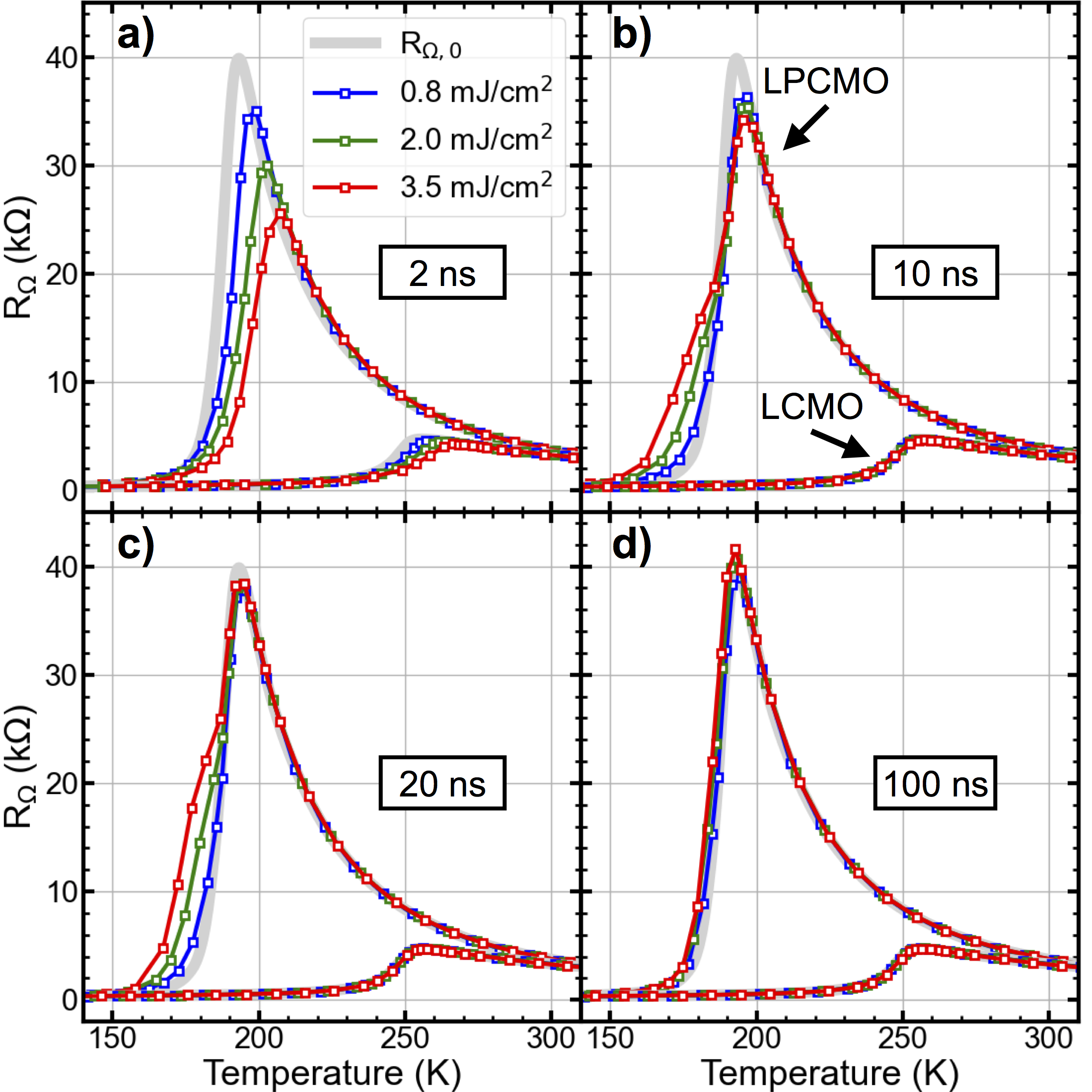}
     \caption{Temperature dependent resistance of LCMO and LPCMO at a) 2ns, b) 10 ns, c) 20 ns, d) 100 ns after laser excitation for three different incident laser fluences highlighting that LCMO returns to the ground state much faster after optical excitation than LPCMO.}
     \label{fig:figure2}
\end{figure}

To gain more insight into the role of nonthermal effects on the resistance dynamics we subtracted the contribution of plain laser heating to the resistance changes for each time step for all measured cryostat temperatures by determining the laser heating in the high-temperature limit (for details see Fig.~\ref{fig:SITemperatureCalc} in the SM). Fig.~\ref{fig:figure2} depicts the T-dependent resistance curves corrected for plain heating for LCMO and LPCMO for different time steps after laser excitation and for three incident laser fluences $F=0.8$, 2.0 and 3.5\,mJ/cm$^2$. Here, transient resistance values which coincide with the equilibrium resistance curve $R_{\Omega,0}$ (gray) correspond to purely thermal physics, any deviations highlight nonthermal contributions to resistance. At $t=2$\,ns after excitation (Fig.~\ref{fig:figure2}a), the maximum laser-induced nonthermal decrease of the resistance has been observed for both samples for all fluences. 10\,ns after excitation (Fig.~\ref{fig:figure2}b) the nonthermal decrease of the resistance is completely recovered in LCMO, indicating that the dynamics of LCMO after this time-delay, shown in Fig.~\ref{fig:figure1}d, is of a purely thermal origin. However, for LPCMO, in addition to the transient metallization, a state of increased resistance appears for $T<T_C$. The recovery of this nonthermal laser-induced state to equilibrium is still not completed after 20 ns (Fig.~\ref{fig:figure2}c) or even after 100 ns (Fig.~\ref{fig:figure2}d). The return to equilibrium finally takes on the order of a microsecond.

\begin{figure}[t]
     \centering
     \includegraphics[width=1\columnwidth]{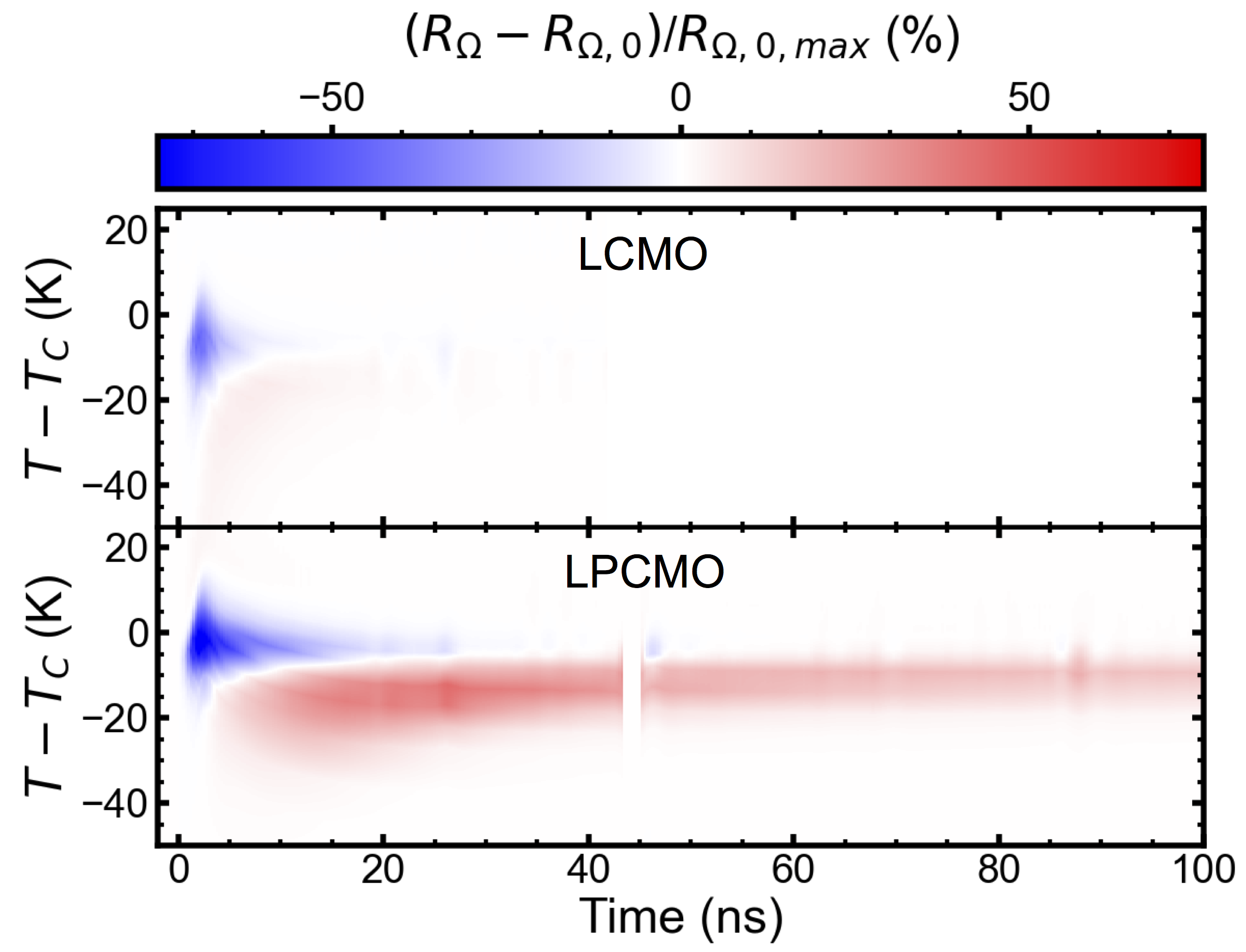}
     \caption{Short-term nonthermal resistance change normalized to the maximum sample resistance in the vicinity of the phase transition for LCMO and LCPMO. Data around 45\,ns was removed due to a measured-setup related artifact, resulting from the signal reflection spike visible in Fig.~\ref{fig:figure1}e.}
     \label{fig:figure3}
\end{figure}

Figure~\ref{fig:figure3} highlights the strong differences in the nonthermal change of resistance between LCMO and LPCMO in more detail. For LCMO, a weak nonthermal decrease of resistance occurs mostly during the presence of the exciting laser pulse and shortly thereafter. In contrast, for LPCMO, a strong decrease in resistance can be seen for timescales an order of magnitude larger than the laser pulse duration, extending up to 20-30 ns closely below the PT. In addition, a long living increase in resistance is visible 10-30 K below the PT in LPCMO, extending to more than 100\,ns time delay.

The origin of these nonthermal resistance changes as well as the differences between LCMO and LPCMO can be rationalized by looking at their static properties crossing the phase boundary. In these materials, an external magnetic field decreases the resistance, yielding the CMR effect (see SM, Fig.~\ref{fig:CPR}b), which is limited to the vicinity of $T_C$, as its origin lies in the phase separated nature of these materials.

In LPCMO, the strong CMR originates from FMM domains which are AFM exchange coupled due to correlated polarons (CPs) located at the domain walls~\cite{Moshnyaga2014}. This AFM coupling breaks down in an external magnetic field, leading to an annihilation of the CPs. As a result, the double exchange interaction and FM alignment is favored leading to an increase (decrease) of electrical conductivity (resistivity). Similarly, we attribute the observed strong nonthermal decrease of the resistance to the laser-induced destruction of CPs~\cite{Fiebig2000, Matsuzaki2009, Wu2009, Caviezel2012, Caviezel2013, Zhang2017} and the following slower recovery of the resistance to the subsequent reformation of the correlated polaronic state. We note that we expect single polarons to reappear almost immediately after annihilation~\cite{Prasankumar2007, LiChung2007, Matsuzaki2009, Wu2009, Zhang2017}. Recovering the CPs, i.e. the initial resistance state, is far slower due to the very large amount of correlated polarons in LPCMO, $n_{CP} \approx 1.5$\,\% of the sample volume required to reconstitute in the presence of strong nanoscale phase separation. Furthermore, due to the mediation of AFM interactions of FM domains by the polarons the recovery dynamics also couples to the magnetic domain structure leading to the observed metastable high resistance state in the FM phase (see Fig.~\ref{fig:figure2}c) discussed further below.

In contrast, LCMO exhibits only $\approx 10^{-3}-10^{-4}$\,\% CPs, which are not able to actuate an AFM exchange coupling between the FM domains. The (uncorrelated) polaronic quasiparticles present in LCMO instead form a disordered gas-like state.  As a result, the CMR effect in LCMO is much weaker than that in LPCMO (see Fig.~\ref{fig:CPR}b in the SM). In perfect agreement to that, the laser-induced annihilation of polarons (and CPs) has less impact in LCMO and due to the disordered character of the system the recovery of the ground state, respectively an equivalently disordered state, proceeds much faster for all temperatures. However, to a certain degree we can achieve a metallization by optical excitation in both material systems - giving rise to an optically induced “colossal photo-resistance” (CPR) effect resembling the CMR. For $F=3.5$\,mJ/cm$^2$ in Fig.~\ref{fig:figure2} this accounts for a $\textrm{CPR}=(R_{\Omega,0} -R_{\Omega,t} )/R_{\Omega,t}\approx 95$\,\% and $\approx 750$\,\% in LCMO and LPCMO, respectively (see SI Fig.~\ref{fig:CPR}). In case of LPCMO, this value is larger than the CMR in an applied magnetic field of $B\approx 1$\,T strength (see Fig.~\ref{fig:CPR}a in the SM).

The transient reflectivity data in Fig.~\ref{fig:figure1}a~and~\ref{fig:figure1}b reflect this different material characteristics as well. The transient metallization in LPCMO appears as a qualitatively visible deviation from thermal behavior due to the so called dynamical spectral weight transfer (DSWT)~\cite{Lobad2000, Lobad2000b, McGill2004, Shen2014}, whereas the influence of nonthermal effects in LCMO is too weak to be seen directly for such long-pulse excitation at this wavelength. The higher sensitivity of transient resistance to the observed metallization processes is easily understood from the fact that resistance measurements probe the states around the Fermi energy, whereas in transient reflectivity all possible optical transitions from initial to final states contribute; depending on the wavelength the sensitivity of transient reflectivity to the PT and to the DSWT will differ.

In addition to the observed transient metallization, a contrasting long-living nonthermal increase of the resistance (and reflectivity) was observed in LPCMO below $T_C$. We attribute this behavior to the optically induced demagnetization creating correlated polarons, which enforce a metastable domain structure, respectively transient phase separation. Laser excitation heats the lattice system with the maximum temperature being reached at about 2-3\,ns. In general, this energy is transferred out of the lattice subsystem by heat diffusion. In case of ferromagnetic ordering, the spin subsystem acts as an additional heat sink for the lattice. Due to the enhanced spin specific heat close to $T_C$, internal equilibration of lattice and spin temperatures transfers a significant amount of the energy stored in the lattice subsystem into the spin subsystem~\cite{Lobad2000, Pincelli2019,Seick2023}. This leads to a quenching of magnetization, i.e., spin disorder on the 2-3 ns timescale, driving the strong initial resistance change below $T_C$ in Fig.~\ref{fig:figure1}e. Afterwards, the thermal gradient leading to thermalization of the spin system with the environment is small, causing a trapping of energy within the spin subsystem. Up to about 20\,ns, an additional increase in resistance is visible for LCPMO in Fig.~\ref{fig:figure1}e for temperatures closely below $T_C$. This is, as discussed before, the timescale on which the polaronic correlations recover after annihilation. This resistance increase below the PT temperature corresponds to the macroscopic establishment of a transient phase separated state with increased AFM-coupling of FM nano-domains due to part of the laser excitation energy still being trapped in the spin system. The emergence of this magnetic structure due to the formation of the correlated polaronic state leads to the observed nonequilibrium increase of the resistance up to about 20 ns. This metastable phase separated state returns very slowly back to the ground state resistance, i.e., the ground state spin structure, as this requires an energetically unfavorable reorientation of AFM aligned FM domains back to the more FM equilibrium configuration at the given base temperature.

\begin{figure}[tbh]
     \centering
     \includegraphics[width=1\columnwidth]{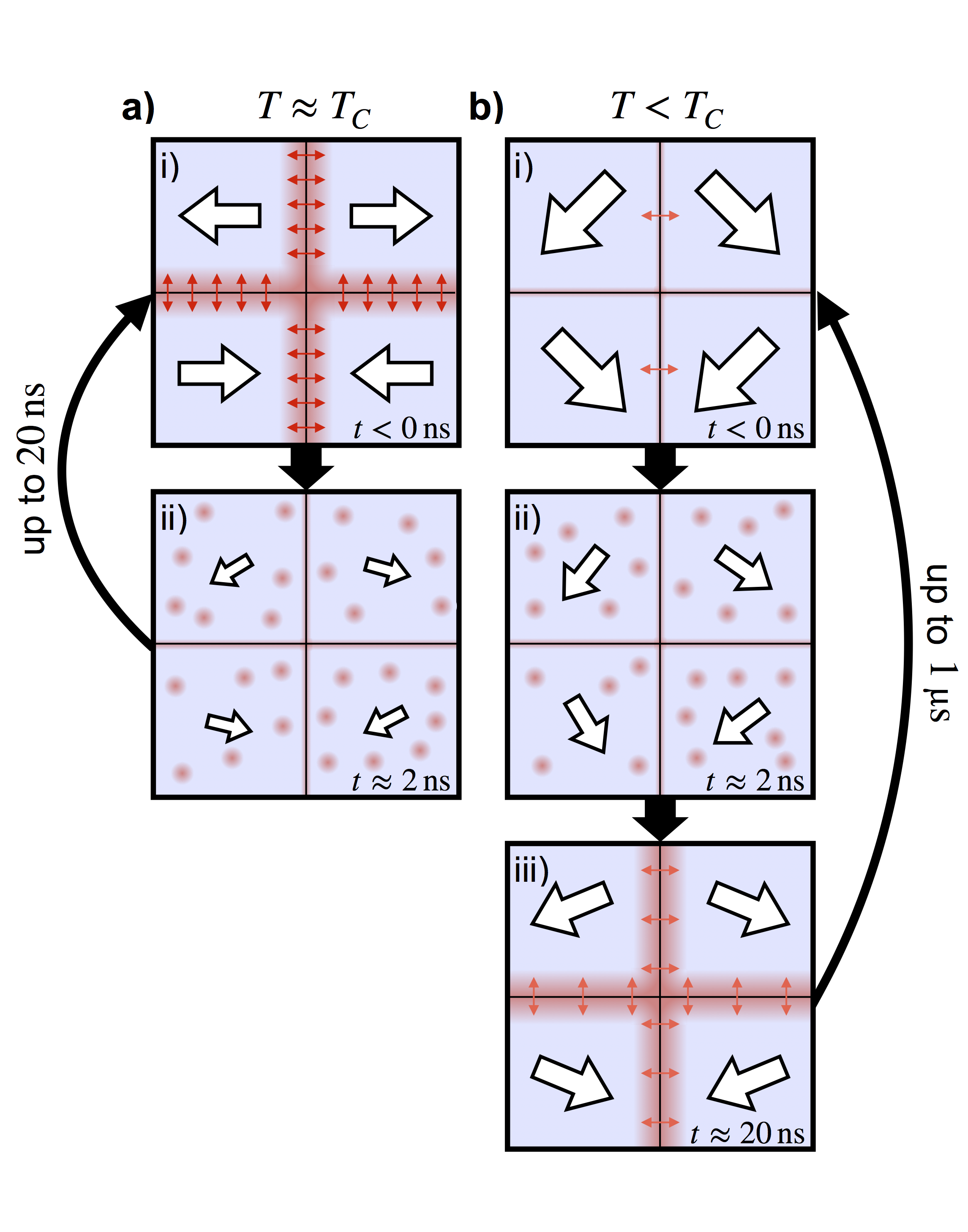}
     \caption{Schematic model to explain the observed dynamics in LPCMO. Arrows designate the macrospin direction within one FM domain, its size indicates the magnetization. For a) $T \approx T_C$, the equilibrium state i) of the system is polaronic insulating with AFM-coupled FM nanodomains. ii) The magnetization is disordered and quenched and correlated polarons are annihilated upon laser excitation decreasing the resistance. Uncorrelated polarons emerge in the FM nanodomains. For b) $T<T_C$, the equilibrium state i) of the system is ferromagnetic and metallic. The magnetization is quenched upon laser excitation increasing the resistance mostly thermally ii) and single polarons emerge in the FM nanodomains. These polarons accumulate at the domain walls iii) and form a correlated state that induces a more AFM alignment of the nanodomains increasing the resistance nonthermally.}
     \label{fig:Model}
\end{figure}


Having discussed both short- and long-term nonthermal resistance changes observed in LPCMO, a global model of the transient phase separated state emerges, as shown schematically in Fig.~\ref{fig:Model} (timescales for $F=3.5$\,mJ/cm$^2$): Initially, at the PT temperature, for vanishing net magnetization, the optical excitation annihilates already existing CPs in LPCMO, leading to the increase in conductivity by the activation of the double exchange mechanism between neighboring domains. Polaronic quasiparticles rapidly reappear afterwards on ps timescales, however, the coalescence of randomly located single polarons into correlated polarons at domain wall boundaries and thus the recovery of the resistance takes place on a much slower 20 ns timescale. Below the phase transition, i.e., still in the FM-ordered state, the situation is slightly different. Lattice and spin temperatures initially rise on similar timescales of 2-3 ns. Energy transfer into the spin system due to the enhanced spin specific heat close to the PT leads to significant nano- to mesoscopic spin disorder with an increased amount of polarons in the system and a concomitant, mostly thermally driven, strong increase in resistance. From this high resistance state, it takes an additional 20 ns to form a new correlated polaronic state actuating additional AFM couplings for the already disordered spin state. This further increases the resistance nonthermally in presence of the already cooling lattice. This state is metastable, as the new spin configuration cannot easily relax back into the equilibrium state expected for the given lattice temperature.

Laser-induced insulator-to-metal transition dynamics were also recently studied by Abreu and coworkers~\cite{Abreu2020}, who compared the recovery of the insulating state after an ultrafast thermal quench into the high-T metallic phase in nickelates, i.e., EuNiO$_3$ (ENO) and NdNiO$_3$ (NNO). ENO, exhibiting a 2$^{nd}$ order PT, shows a fast, exponential recovery of conductivity, while NNO, exhibiting a 1$^{st}$ order PT and electronic/structural phase separation like LPCMO, shows a nonexponential recovery, which is slower than expected from a thermal model. This slower than exponential recovery is believed to be governed by nucleation and growth dynamics. Interestingly, the authors suggest that for NNO the dynamics of electronic and spin degrees of freedom are coupled as well, similar to our discussion on LPCMO. In contrast to Abreu and coworkers, we did observe an influence of phase separation not only in the recovery stage of the system, but also in form of a non-thermal quench into the lower-T metallic state due to the annihilation of (correlated) JT polarons in the manganite systems studied here.

In summary, we explored how the strength of EPS and the amount of correlated polarons influences quasiparticle and spin dynamics after a nanosecond optical stimulus. Close to the metal-to-insulator transition, the quite similar CMR manganites LCMO and LPCMO behave strongly different. For LCMO, with a close to 2$^{nd}$ order phase transition without significant phase separation, we found relatively fast, mostly thermally driven reflectivity and resistance dynamics comparable with the laser pulse duration of 1.25\,ns. This thermal dynamics is superimposed by a weak initial nonthermal metallization of the system due to the annihilation of isolated, uncorrelated JT polarons by the pump excitation. In contrast, LPCMO, with a 1$^{st}$ order phase transition together with significant nanoscale electronic phase separation exhibits a strong and long-lived nonthermal response to the optical excitation. Initially, an almost one order of magnitude stronger resistance decrease was observed due to the annihilation of a large density of correlated JT polarons. This nonthermal state of increased metallicity survives up to 20\,ns, representing the recovery time of the correlated polaronic state. On longer timescales, a nonthermal state of increased resistance forms in LPCMO about 10-20\,K below the phase transition temperature, which persists up to a microsecond. This state is interpreted as a metastable spin-disordered state with frozen magnetic nanodomains that does not correspond to the equilibrium state after the lattice has thermalized by thermal diffusion. Our study demonstrates how the sample nanostructure in form of EPS is crucial for the dynamic evolution of macroscopic sample properties in the otherwise similar CMR materials LCMO and LPCMO. In LPCMO, this manifests itself in the couplings between electrons, lattice and spins mediated by the correlated polarons responsible for the nanoscopic domain structure and allows us to uncover the dynamic evolution of the phase separated state after laser excitation in this material. We expect our results to be transferable to other materials systems, and to provide guidance for the design of materials that can exhibit interesting optically induced metastable and hidden states of matter.

\acknowledgments
T.T.\,,V.M.\,and D.S.\,gratefully acknowledge funding by the DFG, grant no.\,217133147/SFB 1073, project A02.

\bibliographystyle{apsrev4-2}
\include{supplement}
\end{document}

%% file: supplement.tex
\clearpage
\onecolumngrid
\section{Supplemental Material}
\centering\large{Dynamics of electronic phase separation at the laser-induced insulator-metal transition in (La$_{0.6}$Pr$_{0.4}$)$_{0.7}$Ca$_{0.3}$MnO$_3$}\normalsize

\vspace{1em}

by

\vspace{1em}

\centering\large{Tim Titze et al.}\normalsize
\setcounter{page}{1}
\setcounter{figure}{0}
\renewcommand{\figurename}{Fig.}
\renewcommand{\thefigure}{S\arabic{figure}}

\subsection{Samples}
\begin{figure}[h]
     \centering
     \includegraphics[width=0.8\columnwidth]{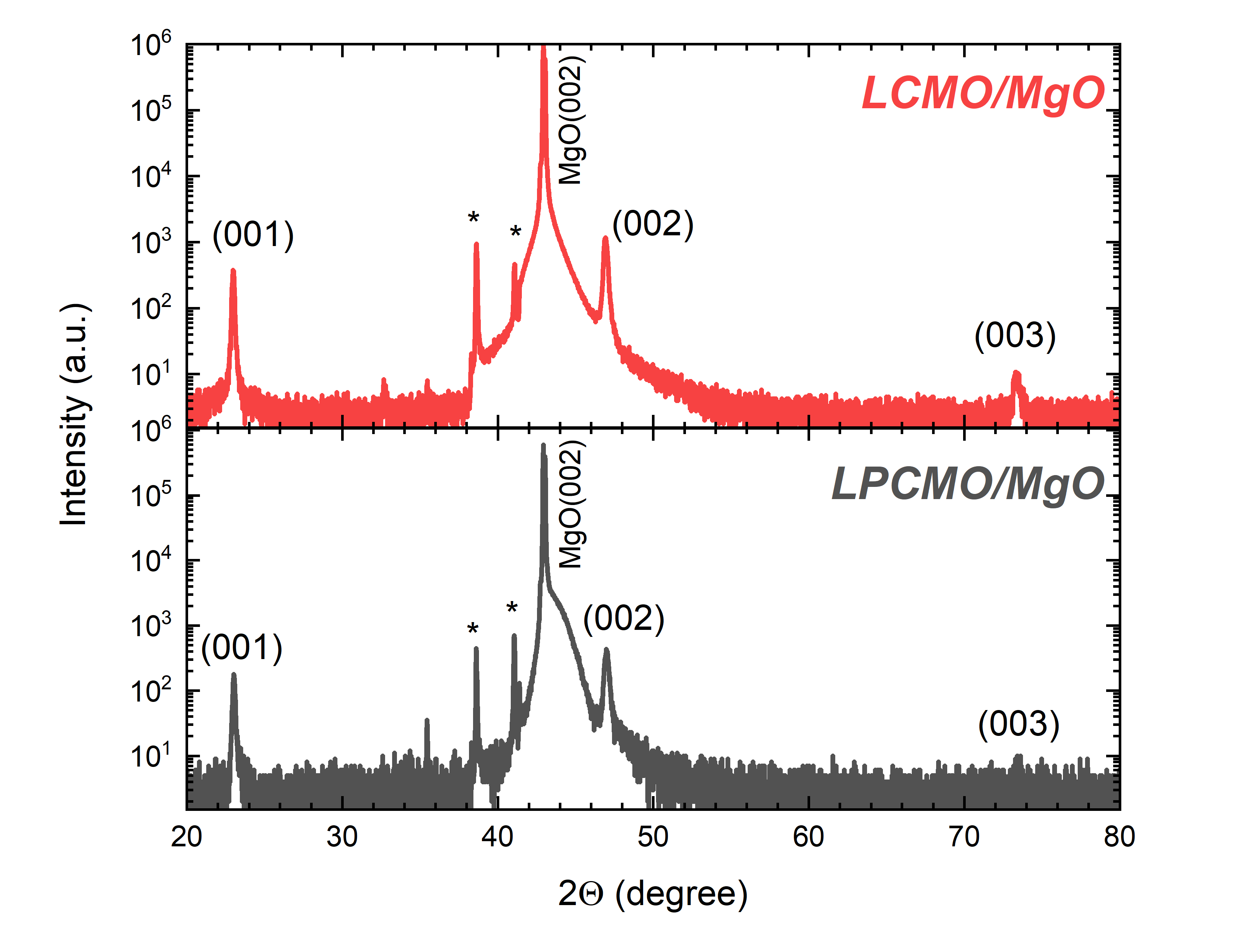}
     \caption{$\Theta-2\Theta$ diffraction patterns of films under study indicates out-of-plane (00l) epitaxy on MgO. Peaks marked with an asterisk are artefacts due to $K_{\beta}$ and impurity lines.}
     \label{fig:SI:XRD}
\end{figure}

Thin films of (La$_{0.6}$Pr$_{0.4}$)$_{0.7}$Ca$_{0.3}$MnO$_3$ (LPCMO) and La$_{0.7}$Ca$_{0.3}$MnO$_3$ (LCMO) with a thicknesses $d=67$ and 56\,nm, respectively, have been heteroepitaxially grown on MgO(200) substrates by means of a metalorganic aerosol deposition technique~\cite{Moshnyaga2014, Seick2023}. The thickness of film samples and their structure have been characterized by X-ray reflection (XRR) and X-ray diffraction (XRD). LCMO and LPCMO exhibit coupled metal-insulator (MI) and a ferromagnet-paramagnet phase transitions at Curie temperature $T_C \approx T_{MI} = 190$\,K and 255\,K, respectively. Being in a nice agreement with the corresponding bulk values~\cite{Uehara1999}, this indicates a stress-free state of films on MgO additionally supported by the closeness of their pseudo-cubic lattice parameters $c_{\mathrm{LCMO}}=0.387$\,nm and $c_{\mathrm{LPCMO}}=0.386$\,nm, determined by XRD (see Fig.~\ref{fig:SI:XRD}) with the corresponding bulk values~\cite{Radaelli1997}. 

\clearpage

\subsection{Experimental Setup}
\begin{figure}[ht!]
     \centering
     \includegraphics[width=0.5\columnwidth]{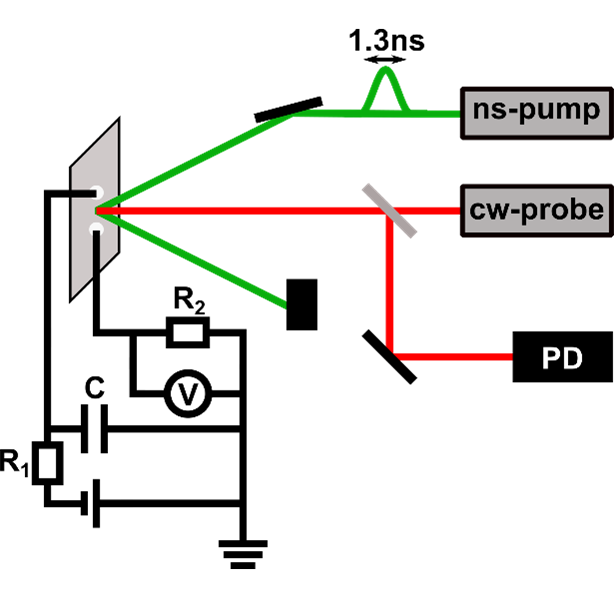}
     \caption{Schematic of the transient reflectivity/resistance setup.}
     \label{fig:SI:Setup}
\end{figure}

The transient reflectivity $\Delta R(t)$ and resistance dynamics $\Delta R_{\Omega}(t)$ were measured simultaneously using the setup shown in Fig.~\ref{fig:SI:Setup}. A pump pulse with duration of 1.25\,ns at a central wavelength $\lambda=515$\,nm has been used to optically excite the sample. The cw-probe (wavelength $\lambda=638.8$\,nm) is focused onto the sample at normal incidence. The reflected probe beam is guided to a fast photodiode (FEMTO HCA-S) connected to an oscilloscope (Agilent Technologies DSO-X 3054A) to measure the pump induced transient reflectivity with time resolution $\sigma_t<2$\,ns. In case of transient resistance, the voltage-drop across the $R_2 = 50$\,$\Omega$ resistor, connected in series with the sample, is monitored by the second channel of the oscilloscope. The resistor $R_1 = 180$\,$\Omega$ and the capacitor $C = 2.2$\,$\mu$F connected in parallel are used to damp voltage oscillations originating from the ultrafast change in sample resistance. The sample has been mounted on a cold finger of a home-made optical cryostat, allowing measurements in the temperature range $T=100-400$\,K using liquid nitrogen as a cooling agent.

\subsection{Full transient reflectivity and transient resistance datasets for LCMO and LPCMO}
\begin{figure}[h]
     \centering
     \includegraphics[width=0.9\columnwidth]{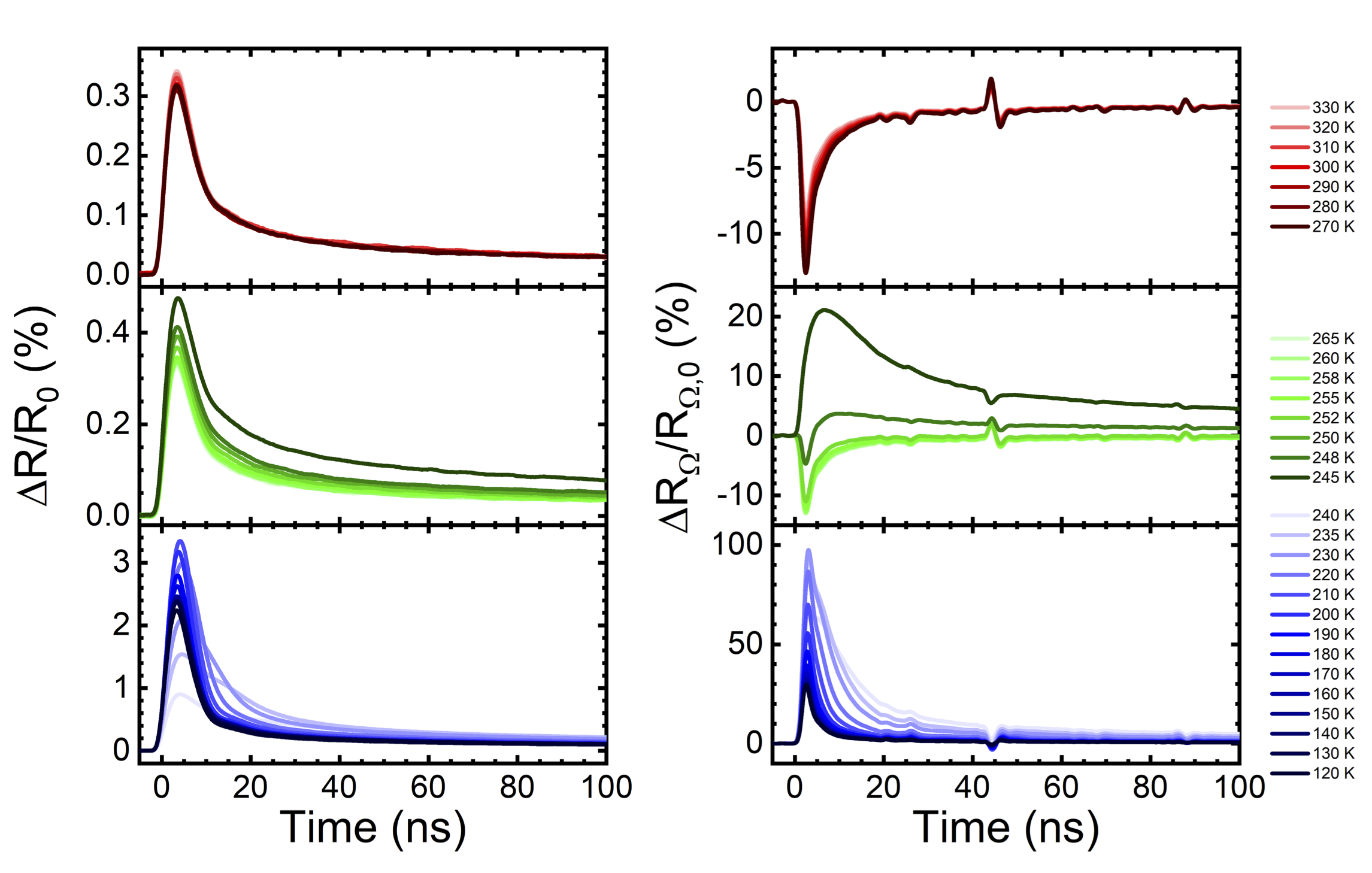}
     \caption{Temperature-dependent change in reflectivity and resistance in LCMO normalized to the values before time zero. Spikes in the resistance data, e.g., at 45\,ns are artifacts from signal reflections due to non-perfect impedance matching.}
     \label{fig:S1}
\end{figure}

\begin{figure}[h]
     \centering
     \includegraphics[width=0.9\columnwidth]{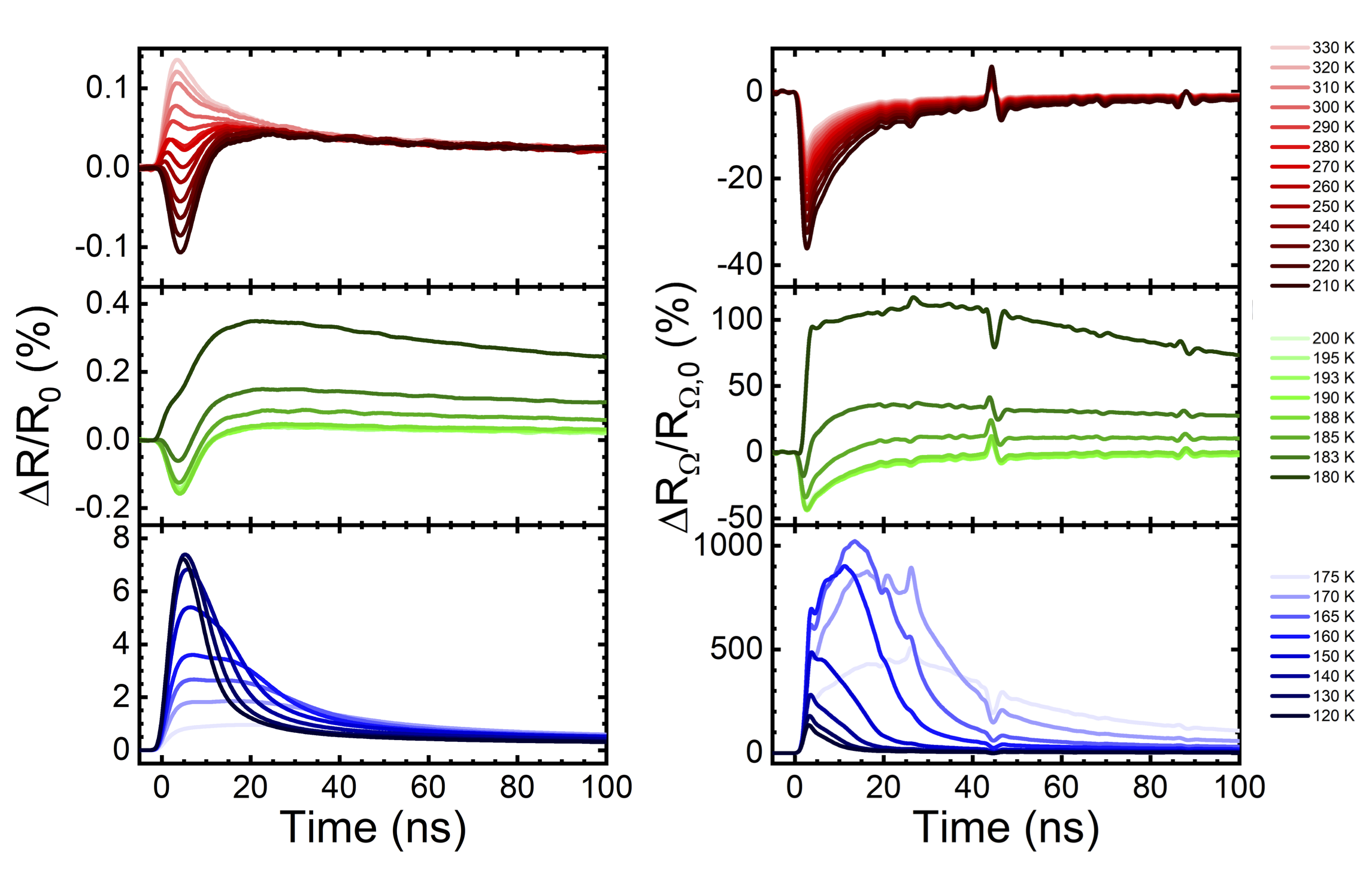}
     \caption{Temperature-dependent change in reflectivity and resistance in LCPMO normalized to the values before time zero. Spikes in the resistance data, e.g., at 45\,ns are artifacts from signal reflections due to non-perfect impedance matching.}
     \label{fig:S2}
\end{figure}

Figures~\ref{fig:S1}~and~\ref{fig:S2} contain the full set of temperature-dependent transient reflectivity $\Delta R/R_0$ and transient resistance $\Delta R_{\Omega}/R_{\Omega,0}$ data measured in the temperature range 120-330 K for LCMO and LPCMO split into two pairs of three sub-figures each. Data within the ferromagnetic metallic phase is given in blue shades, data in the vicinity of the phase transition is given in green shades and data in the paramagnetic insulating phase phase is given in red shades.

\clearpage

\subsection{Extraction of nonthermal dynamics}
\begin{figure}[h]
     \centering
     \includegraphics[width=\columnwidth]{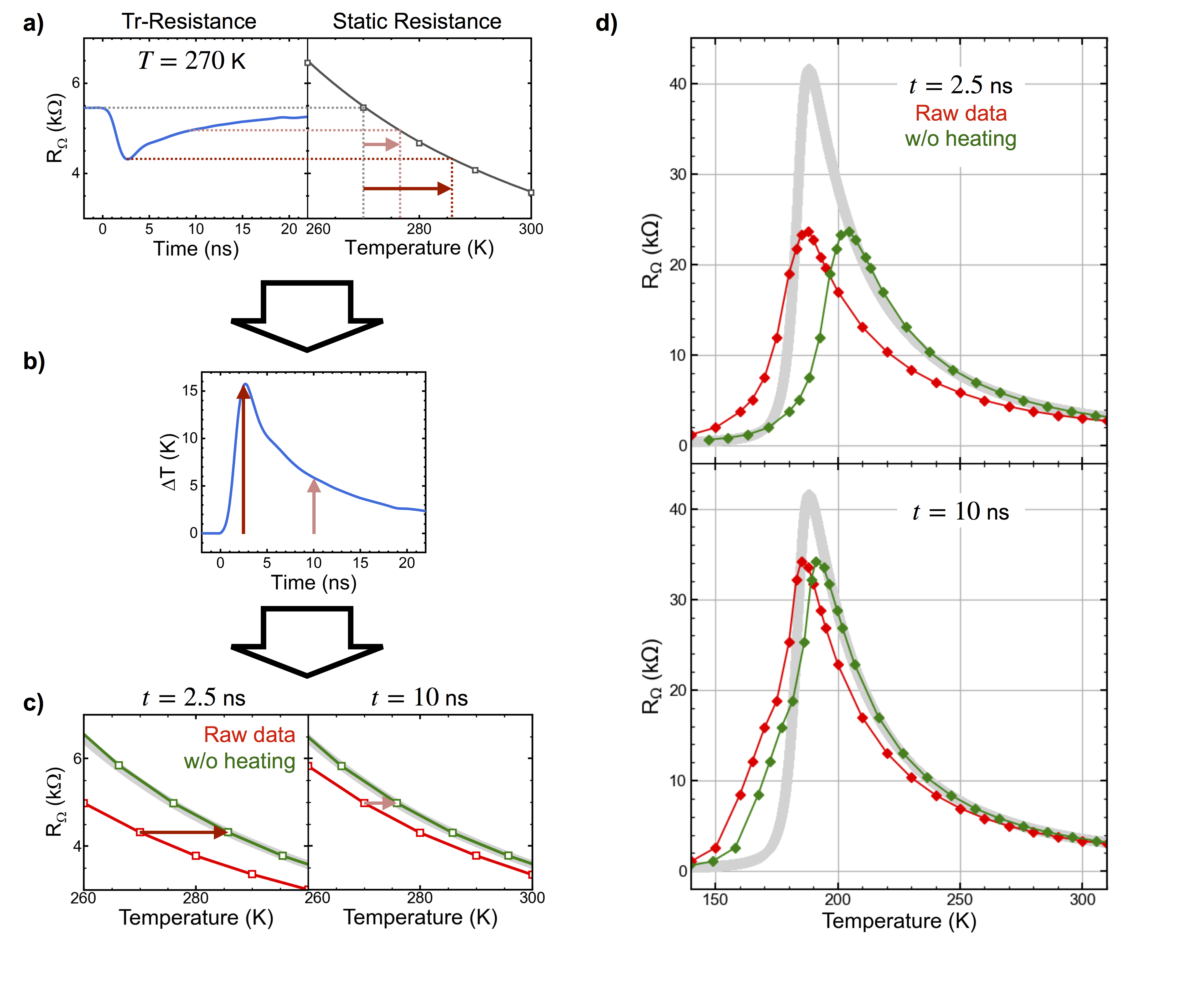}
     \caption{Calculation of the laser induced temperature dynamics. a) The change in resistance for high temperatures ($T\gg T_C$) is mapped onto the interpolated unpumped resistance curve. b) Thus, the temporal evolution of the thermal dynamics is obtained. c) This is then used to correct the $R_{\Omega}(T)$ data for thermal heating at the respective time delays. d) By taking the heat capacity into account, also $R_{\Omega}(T)$ for temperatures close to and below $T_C$ can be corrected.}
     \label{fig:SITemperatureCalc}
\end{figure}

The dynamics can be considered as a pure thermal response for temperatures far above $T_C$. Here, the laser-induced change of resistance can be directly linked to an increase of temperature, which can be extracted as shown in Fig.~\ref{fig:SITemperatureCalc}a in case of LPCMO for 270 K base temperature. We first interpolated the static resistance curves shown in Fig.~\ref{fig:figure1}f. These are then used to map each resistance value in the time resolved data set to a corresponding temperature value, see Fig.~\ref{fig:SITemperatureCalc}a. Thus, we obtain the time-dependent temperature dynamics, which can be depicted in a $\Delta T(t)$ curve for each base temperature, see Fig.~\ref{fig:SITemperatureCalc}b. We then correct the $R_{\Omega}(T)$-data for thermal heating at each time step by shifting each $R_{\Omega}(T)$-data point by the T-increase at the given time step, see Fig.~\ref{fig:SITemperatureCalc}c. For plain sample heating as the reason for the observed resistance changes this then leads to a matching of the equilibrium $R_{\Omega,0}(T)$-data with the $R_{\Omega}(T,t)$ data. As evident from the example dataset shown in Fig.~\ref{fig:SITemperatureCalc}d for LPCMO, this is only true for $T\gg T_C$ at early times after the excitation. In contrast, Fig.~\ref{fig:figure2}b shows this to be nearly perfectly true for LCMO already after $t=10$\,ns.

To extend this procedure from $T>T_C$ to below $T_C$, we used the static heat capacity~\cite{Seick2023} to calculate the increase of the system's energy $\Delta E(t)\propto C \Delta T(t)$, which is assumed to be constant with base temperature (yielding a small error due to T-dependent thermal conductivity and absorption). Considering all base temperatures $T\gg T_C$, we  take the average, yielding the mean energy increase $\Delta \bar{E}(t)$. We can then correct for thermal heating even for temperatures both close to and below $T_C$, as $\Delta T(t)$ is given by $\Delta \bar{E}(t)/C(T)$ (see Fig.~\ref{fig:SITemperatureCalc}d).

\subsection{Colossal Photoresistive Effect and Colossal Magnetoresistance}
\begin{figure}[h]
     \centering
     \includegraphics[width=1\columnwidth]{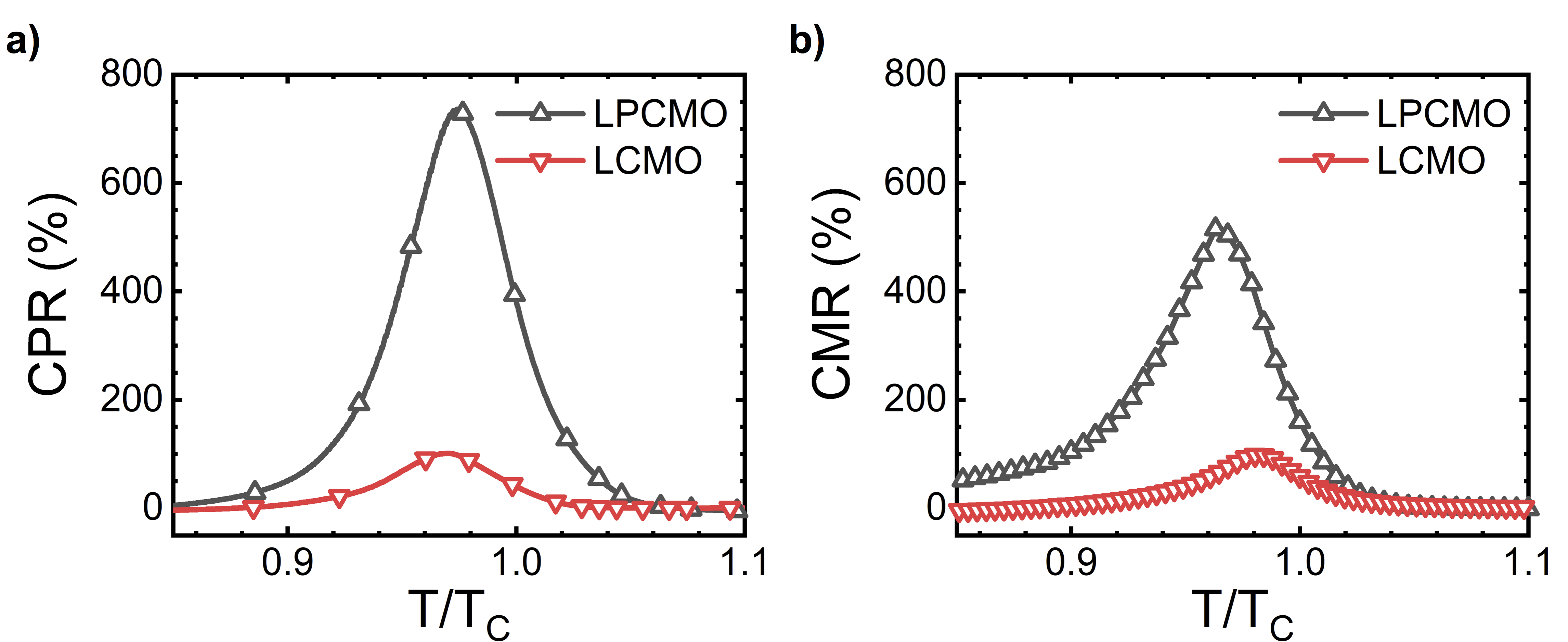}
     \caption{Calculated colossal photoresistance ($t=2$\,ns, F=$3.5$\,mJ/cm$^2$) a) and measured static colossal magnetoresistance ($\mu_0H=1$\,T) b) of LCMO (red downwards pointing triangles) and LPCMO (black upwards pointing triangles).}
     \label{fig:CPR}
\end{figure}

Fig.~\ref{fig:CPR}a showcases the maximum transient enhancement of conductivity in both LCMO (red downwards pointing triangles) and LPCMO (black upwards pointing triangles) calculated from $\textrm{CPR}=(R_{\Omega,0} -R_{\Omega,t} )/R_{\Omega,t}$ at $t=2$\,ns. For LPCMO the maximum conductivity enhancement at about $0.95\cdot T_C$ is nearly eight times higher than for LCMO. As shown in Fig.~\ref{fig:CPR}b, this is comparable to respectivily larger than the colossal magnetoresistive effect in both LCMO and LPCMO, which is induced by an external magnetic field of $\mu_0H=1$\,T strength.
